\documentclass[11pt]{article}
\usepackage[parsep]{collref}
\usepackage{latexsym,amsmath,amssymb,theorem,epsfig}
\usepackage{amsmath}
  
\DeclareFontFamily{OT1}{rsfs10}{}
\DeclareFontShape{OT1}{rsfs10}{m}{n}{ <-> rsfs10 }{}
\DeclareMathAlphabet{\mathscript}{OT1}{rsfs10}{m}{n}

\numberwithin{equation}{section}

\usepackage[T1]{fontenc}
\usepackage[usenames,dvipsnames]{xcolor}
\usepackage[setpagesize=false,pagebackref=false, 
linktocpage, bookmarksopen=true, colorlinks=true, 
linkcolor=blue,citecolor=blue,urlcolor=blue]{hyperref}
\usepackage{tikz} 
\usetikzlibrary{arrows,decorations.pathreplacing,decorations.markings,snakes}
\usetikzlibrary{cd}

\newcommand{\RR}{{\mathbf{\rR}}}

\newcommand{\com}[2]{[#1,#2]}

\def\a{\alpha}

\def\g{\gamma}

\def\z{\psi}

\def\m{\mu}
\def\n{\nu}

\def\s{\sigma}
\def\t{\tau}

\def\z{\zeta}

\def\G{\Gamma}

\def\L{\Lambda}

\def\P{\Pi}

\def\X{\Xi}

\def \S {{\cal S}}

\def\gsim{ \lower .75ex \hbox{$\sim$} \llap{\raise .27ex \hbox{$>$}} }
\def\lsim{ \lower .75ex \hbox{$\sim$} \llap{\raise .27ex \hbox{$<$}} }
\def\be{\begin{equation}}
\def\ee{\end{equation}}
\def\bea{\begin{eqnarray}}
\def\eea{\end{eqnarray}}

\def \ha {{1 \ov 2}}

\def \del{\partial}
\def \a {\alpha}

\def\ov{\over}
\def \ci {\cite}

\def \foot {\footnote}

\def\la{\label}
\def\foot{\footnote}
\newcommand{\rf}[1]{(\ref{#1})}
\def \no {\nonumber}

\def \LL {{\rm L}}\def \z {\zeta}
\def \N {{\cal N}}

\def \g {{\gamma} } \def \hD {\hat \D}
\def \ov {\over}

\def \SS  {{\rm S}}

\def \n {\nu}
\def \N {{\cal N}}
 \def \XX {{\rm X}}

\def \iffa {\iffalse}

\usepackage{tikz}
\usetikzlibrary{decorations.markings}

\makeatletter
\renewcommand*{\@fnsymbol}[1]{\textit{\@alph{#1}}}
\makeatother

\begin{document}
\begin{titlepage}

\title{\vspace{-3cm}
  \hfill{\small Imperial-TP--2025-AT-02  }  
  
   \vspace{1cm} 
   { 
   \bf On   world-volume  supersymmetry \\ 
  of supermembrane  action  in static gauge   }
  \\[1em] }

\author{\large    Arkady A. Tseytlin\footnote{Also at  the Institute  for Theoretical and Mathematical Physics    (ITMP)  of  MSU
      and  Lebedev
    Institute.}  \ \ \ \ \ 
    and \ \ \ \ Zihan Wang\thanks{zihan.wang18@imperial.ac.uk} 
       \\[0.9em]
\it      Abdus Salam Centre for Theoretical Physics \\[0.03em]
  \it   Imperial College London,  SW7 2AZ, UK 
}

  \maketitle

\begin{abstract}
We review  and elaborate on  the issue of  3d world volume supersymmetry 
that appears as a residual part of  global target space   supersymmetry in the   BST supermembrane action.
While  there is no  direct   ``spinning membrane''  analog of the  world-volume 
 supersymmetric spinning   string  action that could be   obtained    by coupling $D$  copies of  3d   scalar multiplet  to 3d supergravity,
   we discuss  how  one  may    construct   an   $\N=1$ 3d  supersymmetric analog of   the derivative expansion of the bosonic  membrane action in  static gauge. We compare  the resulting  $\N=1$   supersymmetric  action for eight  3d   scalar  multiplets 
 to the $\N=8$  3d supersymmetric action   describing  the $D=11$  supermembrane in the static gauge. 
The two actions are not equivalent  which is related to the fact  that 
 the full $\N=8$ supersymmetry  of the  static-gauge $D=11$  supermembrane action can be realised only if  the   fermions  are described  by an $SO(8)$ spinor  rather than  vector.  The two actions   are still directly related  in special  dimensions $D=4$ and 5. 
We  also  compute   the one-loop world-volume     scattering amplitudes for the  two  theories   finding that  they indeed 
agree for $D=4,5$ but   disagree for  $D=11$. 

  \vspace{2cm} 
\hfill{\it Dedicated to the memory of  Kellogg  Stelle}

 \end{abstract}
 
%
\end{titlepage}

\def \iffa  {\iffalse}
\def \RR {{\mathbb R}}
\def \R {{\rm R}}
\def \s {\sigma} \def \t {\tau} 
\def \G {\Gamma} \def \four {{1\ov 4}}
\def \CP  {{\rm CP}}\def \gs {g_s}
\def \hD   {\hat{D}}
\def \JJ {{\rm J}}
\def \te {\textstyle}
\def \zz  {^{(0)}} 
\def \ve {\varepsilon}
\def \zo {^{(1)}}
 \def \rR  {{\rm R}} 
 \def \eps {\epsilon}
 \def \rS  {{\rm S}} \def \rT {{\rm T}}
\def \P   {{\rm P}}
 \def \bbeta {{\rm r}} 
\def \no {\nonumber}
\def \TTT  {{\cal T}} 
\def \S {{\rm S}}  \def \rT  {{\cal T}} 
\def \dDelta {{\hat \Delta}}
\def \ss {\tau}
\def \tn {\lambda} 
\def \hD {\hat D}
\def \half {\tfrac{1}{2}}

\def \ba {\begin{align}}
\def \ea {\end{align}}
\def \RR  {{\mathbb R}}
\def \rr {{\rm R}}  \def \bQ {{\bar Q}}
\def \nf {{\rm n}_{_{\rm F}}}
\def \hT  {\hat T}

 \def \bD {{\rm  b}}  
   \def \rD {{\rm D}} 
\def \LL {{\rm L}}
\def \vt {\vartheta}
\def \ln {{\rm log\,}}
\def \tad {{\rm tad}}
\def \ed {
\small
\bibliography{biblio2.bib}
\bibliographystyle{JHEP-v2.9}
\end{document}
}
\def \edo {\end{document}}

\def \com {} 
\def \un {\underline}
\def \vv {{\rm v}}
\def \umu {\underline \mu}
\def \unu {\underline \nu}
\def \vp  {\varphi }  \def \rr {{\rm r}} 
\def \I {{\mathcal{I}}}  \def \j {{\mathcal{J}}}  
\def \P  {{\cal P}}
\def \vt  {\vartheta}
\def \XX  {{\rm X}} \def \pps {{\Psi}}
\def \X  {{\mathcal{X}}}  \def \SS  {\Sigma} 
\def \ve  {\varepsilon}
\def \L  {{\mathcal  L}}

{\small 
\tableofcontents
}

\section{Introduction}

In view of recent interest in  semiclassical quantization  \ci{Duff:1987cs}
of   11d supermembrane (or M2 brane) 
 in static gauge 
(see, e.g., \ci{Drukker:2020swu,Giombi:2023vzu,Beccaria:2023ujc,Seibold:2023zkz,Beccaria:2023sph,Drukker:2023jxp,Seibold:2024oyr,Giombi:2024itd,Beccaria:2025vdj,
Beccaria:2025npl,Beccaria:2025xry,Gautason:2025per,Gautason:2025plx,Gautason:2025bft,Beccaria:2025ahf})
here we revisit the issue of residual world-volume  supersymmetry  of  its action 
in the case of a flat target space. 


The  target-space supersymmetric   M2-brane action 
 was constructed  as  generalization of the Green-Schwarz \ci{Green:1983wt}
  superstring  action  by Bergshoeff, Sezgin and Townsend    in \cite{Bergshoeff:1987cm,Bergshoeff:1987qx}. The  BST action  has a 
   local $\kappa$-symmetry that ensures  matching of physical bosonic and fermionic degrees of freedom 
 in dimensions $D=4,5,7,11$ \cite{Achucarro:1987nc}. This  matching suggests   the existence 
 of  a hidden 3d  global world-volume  supersymmetry    that should be present in a 
 physical gauge like  static   gauge  \cite{Bergshoeff:1987qx,Achucarro:1988qb}
 (see  also  \cite{Kallosh:1997ky,Kallosh:1997sw,Howe:2004ib,Simon:2011rw}). 

In the   case of the NSR string theory   there is a  spinning string formulation \cite{Brink:1976sc,Deser:1976rb} 
with   manifest    local  world-volume   supersymmetry  that  implies  the existence of    global $\N=1$   
2d supersymmetry in a physical gauge.  
As is well  known, this action  is  effectively   equivalent  
 to the GS  superstring  action  in the light-cone gauge  where  the two  actions become 
 quadratic   in bosons  and fermions  \cite{Witten:1983ymv,Green:2012oqa}  (the  gauge-fixed $D=10$ 
 GS action  has, in fact,  residual $\N=8$  2d supersymmetry that  can only be realised if the fermionic field 
    is  a  space-time spinor).   
It is    less trivial  to show   that   the  two actions are also equivalent   when expanded in powers of derivatives in the static gauge 
\cite{Tseytlin:2025dud}.

In contrast,  there is no  consistent  locally  3d   supersymmetric  ``spinning membrane''  action
that would be equivalent to the BST action  in a physical gauge (see  \cite{Howe:1977hp,Howe:1977us,Howe:1977ut,Bergshoeff:1988ui};
 cf. also \cite{Karlhede:1988xb,Lindstrom:1988az,Luckock:1989jr}).

Instead, one may start with the standard Dirac  action for  the 
 bosonic   membrane,   first   choose a  static gauge 
and  then  $\N=1$, 3d supersymmetrize the  derivative expansion of gauge-fixed action. 

Here we will compare  the resulting world-volume supersymmetric 
 4-derivative  part of  the action   to  its counterpart appearing from 
the BST action  in the static  and appropriate $\kappa$-symmetry gauge   which  also has 
 residual 3d world-volume 
supersymmetry  (extended $\N=8$ one  in the $D=11$ case). 

We will  show that the two actions     can be  put in direct correspondence for  the special 
$D=4$ and $D=5$ cases but disagree for $D=7$ or 11  which has to do 
with the fact that realisation  of $\N >2$ extended 3d  supersymmetry requires the  fermionic field 
(and supersymmetry transformation parameter)  to be a space-time spinor. 
We will check  that the two   actions  lead to different  one-loop 
world-volume   scattering amplitudes for $\hat D=D-3$ physical scalars $X^i$ if $\hat D \not=1,2$.

\

 In section 2 we will  recall  an attempt to construct a  ``spinning membrane''  action   by coupling $D$  copies of  3d scalar  multiplet
 to 3d Einstein   supergravity with a cosmological term. 
In section 3 we will  construct the $\N=1$  3d    supersymmetrization of the quartic  derivative terms in the  expansion of the bosonic membrane   action in static gauge. 
In section 4  we  will review the structure of  the BST action in  static gauge 
and an adapted $\kappa$-symmetry  gauge. 

In section 5   we will   first  explain 
  how the  residual  extended  $\N=D-3$   world-volume 
   3d  supersymmetry appears in the $D$-dimensional 
 BST action after fixing the  static and $\kappa$-symmetry 
 gauges and then compare  the resulting  derivative expansion to the one 
 found in section 3 from the $\N=1$ supersymmetrization of the bosonic membrane action. 
 
In section 6 we  will  discuss  the two corresponding one-loop 
 world-volume S-matrices    showing that they   agree  for   $D=4$  and $5$
 but  disagree  for $D=11$.
 Some  technical details are delegated to  appendices.

\section{3d supersymmetric   membrane    from  coupling to 
   3d supergravity?}

The Dirac   action  for the bosonic membrane   in flat  target space $ \mathbb R^{1, D-1}$  \ci{Dirac:1962iy} 
\begin{equation}
\label{1}
    S=T\int d^3\sigma \,   \L \ ,   \qquad \ \ \L= - \sqrt{-\det (\partial_\m X^A \partial_\n X_A)}\ , \qquad \qquad T= m^3 \ , 
\end{equation}
is classically equivalent to the following analog of the   string action \cite{Brink:1976sc,Deser:1976rb}    containing  an auxiliary 
3d  metric \cite{Howe:1977hp}
\begin{equation}
\label{2} 
    S=-\tfrac{1}{2}  T \int d^3 \sigma \sqrt{-g}\big(g^{\mu \nu} \partial_\mu X^A \partial_\nu X_A-1\big) \ . 
\end{equation}
To  try to find  a  3d  world-volume  supersymmetric version  of \rf{2}  one may follow  the  same idea as in the 2d spinning 
string case
\cite{Brink:1976sc,Deser:1976rb}:   consider $D$ copies  of 
  $\N=1$ 3d scalar multiplets $(X^A, \psi^A)$ and   couple them to  the 3d supergravity fields $(g_{\m\n},\chi_\m)$ 
  \cite{Howe:1977hp,Howe:1977us,Howe:1977ut}.  
  
There are, however,   important differences between 2d and 3d cases:  (i)  in 2d  the pure 
 supergravity terms  are  trivial (total derivatives  which can be dropped)  which is not so   in 3d; (ii) in 2d  there is no cosmological term $\sqrt{-g}$  while 
 in 3d its presence   is required   in order to   have the equivalence  between   \rf{2}   and   the  standard  membrane action  \rf{1}. 
 
 The  corresponding   locally  3d  supersymmetric action may be written as   \cite{Howe:1977ut} 
 (see also \ci{Higashijima:1983wy})\foot{The  3d gamma matrices are chosen to be in real representation and  $\bar{\psi}=\psi^{\mathrm{T}} \gamma^0$ (see  appendix A). 
The covariant derivative is defined as    
$
    \hat D_\mu =\partial _\mu +\frac{1}{4}\hat \omega _{\mu a b}\gamma^{ab}$  where $ \hat \omega_{\mu a b }={\omega}_{\mu a b}(e) +\frac{i}{2}(\bar{\chi}_\mu \gamma_a\chi_b-\bar{\chi}_\mu \gamma_b\chi_a+\bar{\chi}_b \gamma_\mu\chi_a).
$ The 3-bein $e^a_\m$  is related to the  metric by $g_{\m\n}= e^a_\m e^b_\n \eta_{ab}$  and $\gamma_\m =e^a_\mu \gamma_a$.}
 \begin{align}
    \mathcal{S}=\int d^{3}\sigma\, \sqrt{-g} \;\Big\{ M  \Big(R &\te - 
    \frac{i}{\sqrt{-g}}\epsilon^{\mu\nu\rho}\bar{\chi}_{\mu}\hat D_{\nu}\chi_{\rho}
    -\frac{i}{2\sqrt{2}\sqrt{-g}} m \,   \epsilon^{\lambda\mu\rho}\bar{\chi}_{\lambda}\gamma_{\mu}\chi_{\rho}\Big) 
    +\frac{1}{2}m^3 \nonumber\\
    &\te \  +  m^3\Big[ -\frac{1}{2} \partial^{\mu}X^{A}\partial_{\nu}X_{A}-\frac{i}{2}\bar{\psi}^{A}\gamma^{\mu}{\hat D}_{\mu}\psi_{A}+\frac{i}{2} \bar{\chi}_{\mu}\gamma^{\nu}\gamma^{\mu}\psi_{A}\partial_{\nu}X^{A}
    \nonumber\\
    &\te \qquad \quad\  - \frac{1}{16}\bar{\chi}_{\nu}\gamma^{\mu}\gamma^{\nu}\chi_{\mu}\bar{\psi}^{A}\psi_{A}
    +\frac{i}{4\sqrt{2}} M \bar{\psi}^{A}\psi_{A}\Big]  \Big\}\ ,  \label{3}
\end{align}
where $M$   plays the role of  the  3d gravitational constant and $m$ is a mass scale related to the membrane tension in \rf{1}. 
Setting $\chi_\m$, $\psi^A$  and  $M$  to   zero  we have \rf{3}   reducing  to \rf{2}. 
Setting $m=0$ we get   the action of pure 3d supergravity  without a  cosmological term. 

The action \rf{3} is invariant under the following local supersymmetry transformations  ($\ve=\ve (\s)$)
 \begin{align}
&\delta \te X^{A}=im^2\bar{\varepsilon}\psi^{A},\ \ \quad  \qquad\delta\psi^{A}=m^2\gamma^{\mu}(\partial_{\mu}X^{A}-\frac{i}{2}\bar{\chi}_{\mu}\psi^{A})\varepsilon,\nonumber\\
&\te \delta e_{\mu}^{a}=2iM^2\bar{\varepsilon}\gamma^{a}\chi_{\mu},\qquad\quad \delta\chi_{\mu}=4M^2\hat D_{\mu}\varepsilon+\frac{1}{\sqrt{2}}Mm^2\gamma_{\mu}\varepsilon \ . \label{4}\end{align}
The first line in (\ref{3}) is the  3d supergravity action with a  cosmological constant and it  is
separately  invariant under \rf{4}.
 The  last two lines represent the action of  the scalar multiplets coupled 
 to  the supergravity   and  which   is also 
 separately invariant.\foot{The  extra term $\bar\psi^A\psi_A$  term in the  3rd line  in \rf{3} 
 is needed to cancel the $\gamma_\mu \varepsilon$  variation in  $\delta\chi_\mu$ on the second and third line of (\ref{3}).}
  However, if we drop  the   two pure supergravity  terms  in the first line of \rf{3}  
  but keep   the last  cosmological term (as it  is needed to match \rf{2})
   the resulting   action 
   (the    ``spinning  membrane'' action  originally  proposed 
in  \cite{Howe:1977hp})
\begin{align}
    \mathcal{S}'= m^3 \int d^{3}\sigma\,\te  \sqrt{-g} \;\Big\{&\te  -\frac{1}{2} \partial^{\mu}X^{A}\partial_{\nu}X_{A}  -\frac{i}{2}\bar{\psi}^{A}\gamma^{\mu}{\hat D}_{\mu}\psi_{A} + \ha 
    -\frac{i}{2\sqrt{2}\sqrt{-g}} m^{-2}  M \,   \epsilon^{\lambda\mu\rho}\bar{\chi}_{\lambda}\gamma_{\mu}\chi_{\rho} 
     \nonumber\\
    &\  \te  +\frac{i}{2} \bar{\chi}_{\mu}\gamma^{\nu}\gamma^{\mu}\psi_{A}\partial_{\nu}X^{A}
    - \frac{1}{16}\bar{\chi}_{\nu}\gamma^{\mu}\gamma^{\nu}\chi_{\mu}\bar{\psi}^{A}\psi_{A}
    +\frac{i}{4\sqrt{2}} M \bar{\psi}^{A}\psi_{A}  \Big\}\ ,  \label{5}
\end{align}
will not be  supersymmetric  off-shell, i.e. 
   unless   the metric and gravitino are subject to 
differential constraints (equivalent to the supergravity equations following from \rf{3}).

If the  action  \rf{5} were  consistent,  one could  try to  eliminate  $g_{\m\n}$   and $\chi_\m$  algebraically (as
   can be done  in the 2d  case \cite{Brink:1976sc,Deser:1976rb} using also gauge fixing). That would  give 
  a globally supersymmetric action of $X^A$ and $\psi^A$    that  would   represent a
  3d supersymmetric extension of the bosonic  membrane action \rf{1}. 
  However, as was noted in  \cite{Howe:1977hp},  the  action  \rf{5} is invariant under the local   supersymmetry \rf{4} 
   only if  $g_{\m\n}$   and $\chi_\m$   are  subject to the  supergravity field equations that follow  from \rf{3} 
\ba
 \te    \epsilon^{\mu \nu \rho}\big(\mathrm{D}_\nu \chi_\rho+\frac{1}{2\sqrt{2}}m\gamma_\nu \chi_\rho\big)=0 ,
\qquad 
\label{7}
 M\big(R_{\mu \nu}-\frac{1}{2}
g_{\mu \nu}R\big)=\frac{1}{4}m^3 g_{\mu \nu}-\frac{imM}{4\sqrt{2}\sqrt{-g} }\epsilon_{(\mu} \,^{\sigma \rho}\bar\chi_{\nu)} \gamma_\sigma \chi_\rho \ . 
\end{align}
The conclusion is that   one cannot   have a locally sypersymmetric action  that generalizes \rf{2}   in which  $g_{\m\n}$   and $\chi_\m$    appear only algebraically  
 and are unconstrained.\foot{A  no-go theorem was proved in \cite{Bergshoeff:1988ui}: under the assumption that all supersymmetric invariants can be constructed by super-Poincare tensor calculus, there is no  locally supersymmetric
action  that  reduces to  (\ref{1}) when fermions are set to zero and 
the auxiliary metric is integrated out. One attempt  to consider supersymmetrization of  other non-standard  3d  membrane 
actions that are 
not equivalent to  (\ref{2}) but  may reduce to (\ref{1}) when the auxiliary metric is integrated out  but these  will not be discussed  here
(cf. \cite{Karlhede:1988xb,Lindstrom:1988az}).}

Starting   with  the invariant   action \rf{3}  and  solving for   the supergravity  fields would 
 lead to a non-local action for $(X^A,\psi^A)$. One could  contemplate a possibility 
 of considering a  limit of  the two mass parameters $M$ and $m$ in which   non-local  (and higher  derivative) corrections 
 to   the standard membrane action \rf{1}  could be suppressed    but we will not attempt to do  this here. 

\def \DD {{\rm D}}

\section{3d supersymmetrization  of static-gauge membrane action 
   \la{s3}}

Instead of trying to supersymmetrize the reparametrization-invariant  membrane action \rf{1} or \rf{2}   one 
may  use a different approach:
 first fix the   static gauge  $X^\mu=\s^\m$ ($\m=0,1,2$)
 and then introduce 3d superpartners $\psi^i$  for the remaining  $D-3$  
 ``transverse''  coordinate  fields   $X^i$  ($i=1, ..., D-3$). 


In  the  static gauge, the  membrane Lagrangian  in \rf{1} can be expanded as\foot{In  the static gauge we shall use   the  ``flat-space''  indices $a,b,c$  in world-volume derivatives.}  
\begin{align}
\mathcal{L} &\te =\mathcal{L}_2+\mathcal{L}_4 + ...\  , \qquad \ \ \ \  \mathcal{L}_2  =-\frac{1}{2}\del_a X^{i}\del^a X^{i} \ , \la{31} \\
\mathcal{L}_4 &\te =-\frac{1}{8} \del_a X^{i}\del^a X^{i}\del_b X^{j}\del^b X^{j} + \frac{1}{4} 
\del_a X^{i}\del_b X^{i}\del^a X^{j}\del^b X^{j} \ .\label{32}
\end{align}
The  supersymmetric  extension of $\mathcal{L}_2$ is\foot{Compared to \rf{3}
we choose  the sign of the  fermionic  term  to   be   opposite  to that of the bosonic term. 
This  sign   can be  changed by redefining $\g_a$ or flipping the sign of the   the worldvolume coordinates.
}
\ba
\label{33}
& \mathcal{L}^{(0)}=-\tfrac{1}{2}\partial_a X^i \partial^a X^i +i\bar \psi^i \gamma^a \partial_a\psi^i,
\\ 
\label{34}
& \te   \delta ^{(0)}X^i=i\bar\varepsilon\psi^i,\qquad\ \ \  \delta^{(0)} \psi^i =-\tfrac{1}{2}\partial_a X^i \gamma^a \varepsilon\ ,
\end{align}
where  $\g_a$ are    flat-space 3d Dirac matrices and $\big[\delta_1^{(0)}, \delta_2^{(0)}\big] X^i=i\big(\bar{\varepsilon} _1\gamma^a \varepsilon_2\big) \partial_a X^i $. 

To   supersymmetrize  \rf{31}, i.e. $\L \to  \mathcal{L}^{(0)}+\mathcal{L}^{(1)} + ..., $ we  may  deform  the transformations in \rf{34} 
order by order  in derivative (or inverse tension) 
 expansion, $\delta= \delta ^{(0)} + \delta ^{(1)} + ...$, 
requiring $  \delta ^{(0)}\mathcal{L}^{(1)}+\delta ^{(1)}\mathcal{L}^{(0)}=0$, etc.\foot{Similar  perturbative  procedure was used, e.g., in supersymmetrizing the  gauge field $F^2+ F^4 + F^6$ terms in the 
 10d  open string  effective action
 \ci{Metsaev:1987by,Metsaev:1987qp}.}
Since $\delta ^{(1)}\mathcal{L}^{(0)}$  will be proportional to the  leading-order equations of motion 
we can  find $\mathcal{L}^{(1)}$   by demanding its invariance  under  the linearized supersymmetry \rf{34} 
modulo terms  proportional to $\partial_a\partial^a X^i$ and $\gamma^a \partial_a \psi^i$ that can be redefined away 
(and thus  will  not  affect, e.g., the on-shell world-volume  scattering amplitudes). 

We may   then supersymmetrize \rf{32}  by constructing the corresponding 
 superinvariant  in terms of  the 3d superfield 
and covariant derivative  corresponding to \rf{34}\foot{Here $\theta$ is a    3d spinor  Grassmann  coordinate
not to be confused with target space spinor  variable used in later sections.}
\be \la{35}
  \Phi^i=X^i+i\theta \psi^i +\tfrac{i}{2}\theta\theta F^i\ , \qquad \ \ \ \ \ 
  \DD_\alpha=\partial_\alpha-\tfrac{i}{2}\big(\gamma^a \theta\big)_\alpha \partial_a\ .
\ee
Here $\a=1,2$  is the 3d spinor index. 
Since we will be ignoring the terms  proportional  to the leading-order equations of motion, i.e.  interested in  finding  an ``on-shell   superinvariant'',  we may  also set  the auxiliary field $F^i$ to zero. 
The supersymmetry invariants  in general are given   by  $\int d^3\s d^2\theta \, F(\Phi,\DD\Phi, ... )=\int d^3\s \, [F(\Phi,\DD\Phi,  ... )]_{\bar \theta \theta}$. Using that 
\ba
\label{36}
 & \te  \DD_\alpha \Phi^i=i \psi_\alpha^i-\frac{i}{2}\big(\gamma^a \theta\big)_\alpha \partial_a X^i-\frac{1}{4} \theta \theta\big(\gamma^a \partial_a \psi^i\big)_\alpha\, ,
\\
    &\te \DD_\alpha \DD_\beta \Phi^i=-\frac{i}{2} \gamma_{\beta \alpha}^a \partial_a X^i+\frac{1}{2}\big(\gamma^a \theta\big)_\alpha \partial_a \psi_\beta^i-\frac{1}{4}\big(\gamma^a \theta\big)_\alpha\big(\gamma^b \theta\big)_\beta \partial_a \partial_b X^i \ , 
\la{38} \end{align}
we can identify  the  invariants that  have  the  4-derivative terms in \rf{32}  as their bosonic parts:
\begin{align}
\mathcal{I}_1 =&16\Big[\DD^\alpha \Phi^i \DD^\beta \DD^\gamma \Phi^i \DD_\gamma \Phi^j \DD_\beta \DD_\alpha \Phi^j\Big]_{ \bar \theta \theta} =\partial_a X^i \partial^a X^i \partial_b X^j \partial^b X^j-2 \partial_a X^i \partial_b X^i \partial^a X^j \partial^b X^j \nonumber \\
&\qquad \qquad +4 i \partial_a X^j \partial_b X^j\bar{\psi}^i \gamma^b \partial^a \psi^i+4 i \epsilon^{a b c} \partial_a X^i \partial_b X^j\bar{\psi}^i \partial_c \psi^j +2\bar{\psi}^j \gamma_a \partial_b \psi^j\bar{\psi}^i \gamma^b \partial^a \psi^i\ ,\la{39}
\\
\mathcal{I}_2  =&-16\Big[\DD^\alpha \Phi^i \DD^\beta \DD^\gamma \Phi^j \DD_\gamma \Phi^j \DD_\beta \DD_\alpha \Phi^i\Big]_{\bar \theta \theta}=\partial_a X^i \partial^a X^i \partial_b X^j \partial^b X^j\no \\
&\te \qquad \qquad -4 i \partial_a X^j \partial_b X^i\big(\bar{\psi}^j \gamma^a \partial^b \psi^i\big)  +4\bar{\psi}^j \gamma_a \partial_b \psi^j\bar{\psi}^i \gamma^b \partial^a \psi^i-4\bar{\psi}^j \gamma_b \partial_a \psi^j\bar{\psi}^i \gamma^b \partial^a \psi^i\ . \la{40}
\end{align}
Thus the 
 supersymmetrization of the  quadratic and quartic terms in   \rf{31}   is given by\foot{Once again, 
 this expression   is invariant  modulo terms proportional to leading-order  equations of motion that can be redefined away.} 
 \begin{align}
\te   \mathcal{L}=\mathcal{L}^{(0)}-\frac{1}{8}\mathcal{I}_1+ ...
    & =\te -\frac{1}{2}\partial_a X^i \partial^a X^i +i\bar \psi^i \gamma^a \partial_a\psi^i\no \\
   &\te +\frac{1}{4}\del_a X^{i}\del_b X^{i}\del^a X^{j}\del^b X^{j}-\frac{1}{8}\del_a X^{i}\del^a X^{i}\del_b X^{j}\del^b X^{j} \label{41}\\
 &\te  -\frac{i}{2}\del_a X^{i}\del_b X^{i}\bar{\psi}^{j}\gamma^{a}\del^b \psi^{j} -
 \frac{i}{2}\epsilon^{abc}\del_a X^{i}\del_b X^{j}\bar{\psi}^{i}\partial_{c}\psi^{j} -\frac{1}{4}\bar{\psi}^{i}\gamma_{a}\del_b \psi^{i}\bar{\psi}^{j}\gamma^{b}\del^a \psi^{j}+\dots \no
\end{align}
The corresponding   3d action can be 
 dimensionally reduced to 2d  giving a supersymmetric Lagrangian which 
  agrees with the spinning string Lagrangian  of \cite{Brink:1976sc,Deser:1976rb}  when   written 
  in  the static gauge  \cite{Tseytlin:2025dud} (again, up to equation of motion terms).  
  Explicitly,   in 2d we find the following analogs of \rf{39},\rf{40},\rf{41}   (where  now $a,b=0,1$) 
\begin{align}
&\mathcal{I}_1  =\Big[16\DD^\alpha \Phi^i \DD^\beta \DD^\gamma \Phi^i \DD_\gamma \Phi^j \DD_\beta \DD_\alpha\Phi^j\Big]_{\bar \theta \theta}=\partial_a X^i \partial^a X^i \partial_b X^j \partial^b X^j-2 \partial_a X^i \partial_b X^i \partial^a X^j \partial^b X^j\nonumber \\
&\qquad \qquad \qquad \qquad \qquad 
+4 i \partial_a X^i \partial_c X^i\bar{\psi}^j \gamma^a \partial^c \psi^j+2\bar{\psi}^i \gamma_b \partial_a \psi^i\bar{\psi}^j \gamma^b \partial^a \psi^j \ ,  \label{43}
\\
&\mathcal{I}_2 =\Big[-16\DD^\alpha \Phi^i \DD^\beta \DD^\gamma \Phi^j \DD_\gamma \Phi^j \DD_\beta \DD_\alpha \Phi^i\Big]_{ \bar \theta \theta}=\partial_a X^i \partial^a X^i \partial_b X^j \partial^b X^j \no \\ &\qquad \qquad \qquad \qquad \qquad
   -4 i \partial_a X^j \partial_b X^i\bar{\psi}^i \gamma^b \partial^a \psi^j+4\bar{\psi}^i  \partial^a \psi^i\bar{\psi}^j  \partial_a \psi^j \label{44}\ , \\
&   \mathcal{L}=\mathcal{L}^{(0)}-\tfrac{1}{8}\mathcal{I}_1
   =\te -\frac{1}{2}\partial_a X^i \partial^a X^i  +i\bar \psi^i \gamma^a \partial_a\psi^i  \no \\
  & \qquad \qquad \qquad \qquad  +\te \frac{1}{4}\, \partial_a X^i \partial_b X^i \partial^a X^j \partial^b X^j-\frac{1}{8}\partial_a X^i \partial^a X^i \partial_b X^j \partial^b X^j\no \\ 
   &\te  \qquad \qquad \qquad \qquad
    -\frac{i}{2}\,\partial_a X^i \partial_b X^i\bar{\psi}^j \gamma^b \partial^a \psi^j-\frac{1}{4}\bar{\psi}^i \gamma_b \partial_a \psi^i\bar{\psi}^j \gamma^b \partial^a \psi^j+\dots  \ . \label{45}
\end{align}
In Appendix \ref{appB}  we will present the expression for the one-loop  4-scalar scattering amplitude  corresponding to a linear combination of
the  invariants  in \rf{43},\rf{44}. 
 The invariant $\mathcal{I}_1$ can be singled out by requiring integrability, i.e.  that the  tree-level scattering  amplitude 
 satisfies the Yang-Baxter equation  (cf. \cite{Dubovsky:2012sh,Seibold:2024oyr}). 


 Higher order terms can be constructed following similar ``order-by-order'' deformation 
  method  
   by first writing out all possible higher derivative  terms. 
Applying integration by parts, Fierz identity, Majorana properties and dropping terms proportional to equations to motion, the number of possible higher order fermionic  terms is  substantially reduced. Similar to the four-derivative 
 terms  discussed  above, there will be more than one superinvariant as  supersymmetry alone  is not enough to fix the structure 
 of higher-order terms. The additional constraints   that can be imposed are:
 (i)  the bosonic part of the action   should match   the expansion of the Dirac membrane Lagrangian  in the static gauge;
 (ii) the  dimensional  reduction to 2d   should recover  the Lagrangian of the fermionic  string in  the static gauge. 
 
 Higher order $\mathcal N =1$  superinvariants can be constructed  in terms of  superfields using 
  (\ref{36}), (\ref{38}) and similar higher derivative  relations.  One may also try to find a  closed form of the all-order superfield 
  action by analogy  with the 4d   supersymmetric Born-Infeld   action \cite{Cecotti:1986gb,Bagger:1996wp,Rocek:1997hi} (see \ci{Tseytlin:1999dj}  for a review).  The difference compare to 4d case is  that there the action is constructed in terms of chiral superfield
  strength while in 3d one is  to use a real superfield  (cf. also \ci{Hu:2022myf} and a review in \ci{Ivanov:2004zz}). 
  

\section{BST supermembrane action in static  
 gauge}
\label{s4}

Let us now  consider  the   target space 
 supersymmetric  BST  supermembrane  action,    fixing the  static  gauge and  adapted to it $\kappa$-symmetry gauge
 and then expanding in derivatives 
with the aim to compare  to the 3d supersymmetric action  corresponding to  \rf{41}. 
The $D=11$   M2 brane  action of \cite{Bergshoeff:1987qx}  may be written as 
\begin{align}
S_{\rm M} & =S_1+S_2= T \int d^3 \sigma\, L_{\rm M} \ , \qquad \qquad S_1=-T \int d^3 \sigma \sqrt{- g}, \label{51}\\
S_2 & =- \tfrac{i}{2}T \int d^3 \sigma\, \te   \varepsilon^{\mu \nu\rho} \bar{\theta} \Gamma_{MN} \partial_\mu \theta\Big(\Pi_\nu^M\Pi_\rho^N+i \Pi_\nu^M \bar{\theta} \Gamma^N\partial_\rho \theta-\frac{1}{3} \bar{\theta} \Gamma^M\partial_\nu \theta \bar{\theta} \Gamma^N \partial_\rho \theta\Big),\la{52}  \\
g_{\m\n} & =\eta_{MN} \Pi_\mu^M \Pi_\nu^N, \qquad \Pi_\mu^M=\partial_\mu X^M-i \bar{\theta} \Gamma^M \partial_\mu \theta, \qquad \bar{\theta}=\theta^{T} \Gamma^0 \ . \la{53} 
\end{align}
Here  $M,N=0,1,\dots 9, 10$,  and  $\mu ,\nu=0,1,2$  and we assume  a real representation of  $\Gamma^M$  Dirac matrices. 
In addition to the standard   bosonic symmetries  the  action is invariant  under global 11d supersymmetry  with
constant  parameter $\epsilon$ and 
local $\kappa$-symmetry   with parameter $\kappa(\s)$ 
\begin{align}
& \delta X^M=i\bar{\theta} \Gamma^M
(1+\Gamma)\kappa+i\bar{\epsilon} \Gamma^M \theta \ , \qquad \qquad
 \delta\theta=(\mathbf{1}+\Gamma) \kappa+\epsilon\ ,\label{54} \\ 
&\Gamma  \equiv \te \frac{1}{3!\sqrt{-g}} \epsilon^{\mu \nu\rho} \Pi_\mu^M \Pi_\nu^N \Pi_\rho^P \Gamma_{MNP} \ , \qquad \ \ \ \ \ 
\Gamma^2=1,\qquad  \text{Tr}\,\Gamma=0 \ . \la{55}
\end{align}
Fixing the static gauge $X^\mu=\sigma ^\mu$  we find that $\Gamma$ in \rf{55} expanded in  powers of derivatives of 
the transverse coordinates  $X^i$    becomes  ($a=0,1,2;$  $i=3, ...,10$) 
\begin{equation}
       \Gamma=\gamma- \partial_a X^i \, \Gamma^a \Gamma_i \gamma \, +\dots,\qquad\qquad \  \gamma\equiv \Gamma^{012} \ , \qquad \gamma^2 =1 \ . \la{56}
\end{equation}
Let us split $\theta$  and the parameter $\kappa$  into parts with definite $\gamma$ chirality
\begin{align}
    &\theta=\P_-\theta + \P_+ \theta \equiv \vartheta+\tilde\vartheta,\quad\qquad \ \ \ 
    \kappa=\P_+ \kappa + \P_- \kappa \equiv \varkappa+\tilde\varkappa\ ,\label{58}\\
  &\qquad \ \      \mathcal{P}_\pm\equiv \tfrac{1}{2}(1\pm\gamma),\qquad\qquad 
      \mathcal{P}_\pm ^2=\mathcal{P}_\pm\ . \la{57}
\end{align}
In the static gauge   where $\Gamma=\gamma + ...$ in  \rf{56}
 we   can   fix the  $\kappa$-symmetry   gauge as\footnote{An equivalent gauge  was   considered  in \cite{Bergshoeff:1987qx} using  $\tilde{\mathcal{P}}_\pm=\frac{1}{2}(1\pm\tilde\gamma)$ where  $\tilde \gamma\equiv \Gamma^{34... 910}$ (see also \ci{Bergshoeff:1987dh}).}
\begin{equation}
\label{60}
    \tilde \vartheta=\mathcal{P}_+\theta=0 \ , \ \ \ \   {\rm i.e.} \ \ \ \      \theta= \vartheta \ . 
\end{equation}
Note that in the  static  gauge only $ \varkappa=\mathcal{P}_+\kappa$  is involved in the  linearised  transformations in \rf{54}
so we may set $ \tilde \varkappa=\mathcal{P}_-\kappa=0$. 

The BST  Lagrangian in \rf{51}   in the   static gauge and  the $\kappa$-symmetry gauge \rf{60}  and 
expanded in powers of derivatives  can be written as  \cite{Seibold:2024oyr,Beccaria:2025xry}\foot{Here $a,b,...=0,1,2$ are contracted with $\eta_{ab}$.
We rescaled  $\vartheta\rightarrow \frac{1}{\sqrt{2}}\vartheta$  to make the fermionic  kinetic term  normalized as in \rf{33}.}
    \begin{align}
    \label{61}
L _{\rm M}  =&\te -\frac{1}{2} \partial_a X^i \partial^a X^i+i \bar{\vartheta} \Gamma^a \partial_a \vartheta\nonumber \\
&\te  +\frac{1}{4} \partial_a X^i \partial_b X^i \partial^b X^j \partial^a X^j-\frac{1}{8} \partial_a X^i \partial^a X^i \partial_b X^j \partial^b X^j \nonumber\\
& \te -\frac{i}{2} \partial_a X^i \partial_b X^i \bar{\vartheta} \Gamma^a \partial^b \vartheta
-\frac{i}{4} \epsilon^{a b c} \partial_a X^i \partial_b X^j \bar{\vartheta} \Gamma_{i j} \partial_c \vartheta  -\frac{1}{4} \bar{\vartheta} \Gamma_a \partial_b \vartheta \bar{\vartheta} \Gamma^b \partial^a \vartheta+\ldots\ .
\end{align}
The  double dimensional  reduction of the  action \rf{51}   gives \ci{Duff:1987bx} 
the type IIA GS  string action 
\begin{align}
S_{\rm IIA}  =S_1+S_2=-T \int d^2 \sigma \sqrt{- g} -iT \int d^2 \sigma\;  \varepsilon^{\mu\nu} \bar{\theta} \Gamma_M \Gamma_{11} \partial_\mu \theta\big(\Pi_\nu^M+\tfrac{i}{2} \bar{\theta} \Gamma^M \partial_\nu \theta\big)\ , \la{62}
\end{align}
where $\theta$  may be split into  two 10d Majorana-Weyl fermions $(\theta^1,\theta^2) $ with opposite chirality.
This  action is invariant under 
\begin{align}
& \delta X^M=i\bar{\theta} \Gamma^M (1+\hat \Gamma\Gamma_{11})\kappa+i\bar{\epsilon} \Gamma^M \theta,\qquad\qquad  \delta \theta=(\mathbf{1}+\hat \Gamma\Gamma_{11}) \kappa+\epsilon,\label{63}
\\
&\hat \Gamma\te   =\frac{1}{2!\sqrt{-g}} \epsilon^{\mu \nu} \Pi_\mu^M \Pi_\nu^N\Gamma_{MN},\quad\qquad\qquad \qquad \ \ \ \  \Gamma_{11}\equiv \Gamma^{012...89}\ . 
\la{64} \end{align}
In the   static gauge the $\kappa$-symmetry  gauge is fixed as 
\be  \mathcal{P}_+\theta=0 \ , \qquad 
\ \ \ 
   \mathcal {{P}}_\pm\equiv \tfrac{1}{2}(1\pm \gamma \Gamma_{11}),\quad\ \ \ \  \mathcal {P}_\pm^2=\mathcal{P}_\pm,\quad\ \ \ \  \gamma \equiv \Gamma^{01}.\label{65}
\ee
The derivative  expansion of the  GS string Lagrangian following from \rf{62} 
then  takes the same form as in (\ref{61})  but without the  $\epsilon^{abc}$ term, i.e. 
\begin{align}
L _{\rm IIA}=&\te -\frac{1}{2} \partial_a X^i \partial^a X^i+i \bar{\vartheta} \Gamma^a \partial_a \vartheta\nonumber \\
&\te  +\frac{1}{4} \partial_a X^i \partial_b X^i \partial^b X^j \partial^a X^j-\frac{1}{8} \partial_a X^i \partial^a X^i \partial_b X^j \partial^b X^j \nonumber\\
&\te  \label{66}-\frac{i}{2} \partial_a X^i \partial_b X^i \bar{\vartheta} \Gamma^a \partial^b \vartheta -\frac{1}{4} \bar{\vartheta} \Gamma_a \partial_b \vartheta \bar{\vartheta} \Gamma^b \partial^a \vartheta+\dots \ . 
\end{align}

\section{Worldvolume supersymmetry of   static-gauge  BST   action\label{s5}}

Before gauge fixing, the BST  action in flat  target space background  is invariant  under global 11d super Poincare  symmetry and 
 local  reparametrizations and $\kappa$-symmetry.  Choosing    static gauge  breaks the  global 
 bosonic  symmetry $ISO(1,10)$  to $ISO(1,2)\times SO(8)$.
 Fixing the  $\kappa$-symmetry gauge \rf{60} leaves  a  part  of the   global spacetime supersymmetry 
 that may be interpreted as  a worldvolume supersymmetry  \cite{Bergshoeff:1987qx,Achucarro:1988qb,Kallosh:1997ky,Kallosh:1997sw,Howe:2004ib,Simon:2011rw}.

Requiring the residual symmetry transformations to preserve the gauge choices   gives conditions on 
the parameters of  the 3d diffeomorphisms   and  the $\kappa$-symmetry. 
The preservation of the  static gauge $X^a=\s^a$ implies 
$
     \delta X^a=\zeta^b\partial_bX^a+i\bar{\theta} \Gamma^a (1+\Gamma)\kappa+i\bar{\epsilon} \Gamma^a \theta=0,
$
which   gives  $\zeta^a=-i\bar \theta \Gamma^a (1+\Gamma)\kappa -i\bar \epsilon\Gamma^a \theta$. Then the 
 combined    reparametrization   and symmetry transformations  in \rf{54} take the form 
\begin{align}
    &\label{rs11}\delta X^i =\big[-i\bar \theta \Gamma^a (1+\Gamma)\kappa  -i\bar \epsilon\Gamma^a \theta\big]
    \partial_a X^i+i\bar \theta \Gamma^i (1+\Gamma)\kappa +i\bar \epsilon\Gamma^i \theta ,\\
    &\label{rs21}\delta \theta=\big[-i\bar \theta \Gamma^a(1+\Gamma)\kappa -i\bar \epsilon\Gamma^a \theta\big]\partial_a\theta +(1+\Gamma )\kappa +\epsilon \ , 
\end{align}
with the linear part   being (cf. \rf{56}) 
\begin{align}
    &\label{rs1}\delta X^i =i\bar \theta \Gamma^i (1+\gamma)\kappa +i\bar \epsilon\Gamma^i \theta + ... \, ,\ \ \ 
    \qquad \ \ \ \ \delta \theta=(1+\gamma )\kappa +\epsilon + ... \ . 
\end{align}
Preserving the $\kappa$-symmetry gauge (\ref{60})  implies  (using definitions in   \rf{58})
\begin{align}
\delta\tilde\vartheta&=
    2\varkappa+ \gamma\partial_a X^i \Gamma^a \Gamma_i \tilde \varkappa+\xi=0\ ,\label{kappac} \ \quad   {i.e.} \quad \ \ 
     \varkappa=-\tfrac{1}{2}\xi+\dots \ , 
    \qquad \quad \epsilon = \P_+ \epsilon + \P_- \epsilon \equiv \xi + \tilde \xi \ .
\end{align}
The residual global supersymmetry of  the gauge fixed  BST action is then 
\begin{align}
     \label{67}   \delta X^i =i\bar{\xi}\Gamma^i\vartheta  +\dots
\ ,  &\qquad  \qquad \delta\vartheta =\tfrac{1}{2}\partial_a X^i\Gamma^a \Gamma^i\xi+\dots
  \ , \\ &  \delta \vt = \tilde \xi \ .  \label{677} 
\end{align}
Here  the target-space  spinor $\xi$  may be interpreted  is a  parameter  of the 
 $\N=8$  3d   global   world-volume   supersymmetry  and 
 $\tilde\xi$  as  a  parameter  of the  shift  (``goldstino'') symmetry  complementing  the bosonic   shift 
   symmetry  $\delta X^i =  a^i$ of the action \rf{61}.  
   In what follows we will focus on the  world volume   supersymmetry transformations in  \rf{67}.

While the world-volume S-matrix    for scattering of $(X^i, \vt)$   will be invariant under the  linearized  3d supersymmetry in \rf{67}
(which is obviously  the symmetry of the quadratic terms in \rf{61})\foot{As usual, non-linear  terms in the symmetry  transformations do not  contribute when  considering the transformations  acting on on-shell scattering amplitudes. 
Note also that  
   the    3d supersymmetry algebra  corresponding to \rf{67}  (i.e.  $  \big[\delta_1, \delta_2\big] X^i=i\big(\bar{\xi}_1  \Gamma^a \xi_2\big) \partial_a X^i+...$, etc.) 
   closes on $\vt$ only on-shell.}
the 3d supersymmetry invariance of the off-shell   action  corresponding to \rf{61} 
will require also   to include  non-linear terms  denoted by dots in \rf{67}. 

Restricting attention to quadratic and quartic terms in \rf{61}   we may ignore the  non-linear modification  of \rf{67}
as the  variation of the  quadratic terms   will then be proportional  to  the  free equations of motion.
Thus  the quartic terms  in \rf{61}  should   be invariant under the linearized transformations in \rf{67} modulo equation-of-motion terms  that can be redefined away. Thus   these quartic terms   should 
represent an on-shell   invariant of the   world-volume supersymmetry in \rf{67}.

 In general the $\N=1$   supersymmetric Lagrangian   \rf{41}  is   defined  for any  number $\hat D = D-3 $ of $(X^i,\psi^i)$ multiplets, 
 but  the  BST action is $\kappa$-symmetric  only for  $D=4,5,7,11$   when   the static-gauge  Lagrangian 
  \rf{61}    has $\N=\hat D = 1,2,4,8$   extended  3d supersymmetry. We will  first focus  on $D=11$ case and  then comment on other
  special  cases.

\subsection{Comparing $\N=1$  and $\N=8$ supersymmetric actions}

Let us   now  compare the  two  Lagrangians   in \rf{41}  and \rf{61} that   have the same bosonic parts  depending on $X^i$ 
while  the fermions  being  $8$   Majorana  3d  spinors $\psi^i$ (with  total of $8\times 2$   real components)  in \rf{41} 
 and 
one 11d Majorana spinor $\vt$   (subject  to $\P_+ \vt=0$, i.e. with  16 real independent components) in \rf{61}. 

The 
Lagrangian in \rf{41} is  invariant under  the standard $\N=1$ 3d supersymmetry \rf{34} parametrized  by  one  constant 
    3d spinor $\ve$   with 2 real components while 
the one in  \rf{61}  is  invariant under the $\N=8$ 3d supersymmetry \rf{67}  parametrized  one constant   11d   target space spinor $\xi$  
(subject to $\P_- \xi=0$, i.e.    with 16 independent   real components).\foot{For the appearance  of $\N=8$  3d supersymmetry
 in super-embedding 
  approach see \ci{Bandos:1995zw,Sorokin:1999jx,Bandos:2023web}  and also \ci{Howe:2004ib}; 
   for discussions  of  $\N=8$  supersymmetry  in other  contexts see, e.g.,  \cite{Marcus:1983hb,Samtleben:2009ts}.}

To compare the  two actions  and their supersymmetries 
let us  decompose the  11d   Dirac  matrices into 3+8 parts as  
\begin{align}\label{68}
 &   \Gamma^a=\gamma^a \otimes \rho ^*, \qquad \Gamma^i=\mathbb{I}\otimes \rho^i, \qquad 
    \gamma=\Gamma^{012}=-\mathbb{I}\otimes \rho^*, \qquad   \ \   \{\rho_i,\rho_j\}=\delta_{ij} \ , 
\\
 &\la{69} 
  \mathcal{P}_\pm=\half ( 1 \pm \g) = \mathbb{I}\otimes p_\mp,\qquad\quad\ \  \ \ p_\mp=\tfrac{1}{2}(1\mp\rho^*)\  , \ \ \ \ \ \ \ 
  \qquad \rho^*=\rho^{3456789 10}\ .
\end{align}
Similarly, let us decompose the target space  spinor field  $\vartheta(\s)$
and  the global supersymmetry parameter $\xi$   as
\begin{equation}\label{70}
 \vartheta (\s)  =\psi^i(\s)  \otimes \eta^i\  , \qquad  \ \ \ \ 
    \xi=\varepsilon\otimes\mathcal{X} ,\qquad\ \ \ p_-\eta^i =0  \ , \ \  \ \ \ 
    p_+\mathcal{X} =0 \ .
\end{equation}
Here 
 $\psi^i(\s) $  are 8 anticommuting    2-component   fields, 
 $\varepsilon$ is a constant anticommuting   2-component  spinor,   while 
 $\eta^i$  and $\X$   are   constant  commuting 8d spinors with   8  independent real components each. 
  Since $\psi^i(\s)$  represent  the same number of  independent functions  as the 16-component $\vt(\s)$ 
  it may be possible to choose $\eta^i$  in the special  form 
  \be \la{700}
   \eta^i =  \rho^i\SS \ , \qquad \ \  {\rm i.e.} \ \ \ \ \   \vartheta (\s)  =\psi^i(\s)  \otimes   \rho^i\Sigma \ ,   \ee
   where $\Sigma$ is a constant  ``reference''    8d spinor. 
   If we further specify the  parameter $\xi$ in \rf{70}   so that its 8d spinor part  $\X$ is  fixed as 
   \be \la{701} 
   \X= - \SS \ , \qquad \ \  {\rm i.e.} \ \ \ \ \   \ \ \ \ \xi=- \varepsilon\otimes\SS\ , \ee
   then 
    the linearized   supersymmetry transformations in \rf{67}  may be written as (cf. \rf{34}) 
\begin{align}
&\delta X^i =\lambda^{ij}  \delta^{(0)} X^j \, , \qquad 
 \delta \vartheta= \delta^{(0)} \psi^i   \otimes  \rho^i\SS  \ , \qquad \ \ \ 
  \lambda^{ij}  \equiv  \SS^T\rho^i\rho^j \SS = \delta^{ij}  \SS^T \SS \ ,  \la{71} \\
& \delta^{(0)} X^i =  i\bar\varepsilon\psi^i\,,\ \ \qquad \ \ \   \delta^{(0)} \psi^i  = - \tfrac{1}{2}\partial_a X^i \gamma^a\varepsilon \ . \la{72}
\end{align}
Thus   for this special choice of the parameter $\xi$  we  get 
  the $\N=1$  supersymmetry transformation in \rf{34} or \rf{72}  
 as a  special  case of the $\N=8$ supersymmetry transformation in \rf{67}.
  

  The Lagrangians   in \rf{41}  and \rf{61}  have  the same   structure of terms involving $\G^a$   matrices  (which can be 
  identified with $\g^a$ as in \rf{68})   but  differ in  $\epsilon^{abc}$ terms. 
   Using  the identification in \rf{700}   we find 
  \begin{align}
\bar{\vartheta}\Gamma^a\partial_b\vartheta&\ \ \to \ \  V \bar \psi^i\gamma^a\partial_b\psi^i     \ , \qquad\qquad \ \ \  \qquad \qquad  \ \ V\equiv \SS^T\SS ,\la{74} \\
  \bar{\vartheta} \Gamma^{ij}\partial_c \vartheta    &\ \ \to \ \  2V \bar\psi^{[i}\partial_c\psi^{j]}
  + U^{ijkl} \bar\psi^k\partial_c\psi^l\ , \qquad\ \ \ 
  U^{ijkl} \equiv  \SS^{T}  \rho^{ijkl}\SS\ , \label{75}
\end{align}
where $V$ is a constant   scalar and    $U^{ijkl} $   is a  totally antisymmetric constant $SO(8)$ tensor. 
 $V$ can  thus  be  absorbed into a rescaling of $\psi^i$  and 
then  all but the  $\eps^{abc}$ terms  in \rf{41}  and \rf{61}    become equivalent. 

The  $\frac{i}{4} \epsilon^{a b c} \partial_a X^i \partial_b X^j \bar{\vartheta} \Gamma_{i j} \partial_c \vartheta$  term 
in \rf{61}   would   also  match the $ \frac{i}{2}\epsilon^{abc}\del_a X^{i}\del_b X^{j}\bar{\psi}^{i}\partial_{c}\psi^{j}$ 
term in \rf{41} if not for the  second $U^{ijkl} \bar\psi^k\partial_c\psi^l$     structure  in \rf{75}. 
This implies   that 
 that the   actions  \rf{41}  and \rf{67}  are not, in general,   equivalent. 
Indeed, as we will check  below,  the  contributions to the one-loop 
 2-scalar scattering amplitude coming from the $\epsilon^{abc}$ terms in   \rf{41}  and \rf{61}    differ by a factor of 4, 
so that  the two S-matrices are not the same.

 One may wonder how this  inequivalence of  \rf{41}  and \rf{61}   can be consistent with the fact that 
  both actions are  invariant under the $\N=1$ 3d supersymmetry (cf. \rf{71},\rf{72}). 
   As we will show in appendix B, 
 the  extra $U^{ijkl}$-dependent  term in  \rf{61}
 originating from $\frac{i}{4} \epsilon^{a b c} \partial_a X^i \partial_b X^j \bar{\vartheta} \Gamma_{i j} \partial_c \vartheta$
 upon  using  \rf{75}, i.e. $\frac{i}{4} \epsilon^{a b c}  U^{ijkl} \, \partial_a X^i \partial_b X^j \, \bar\psi^k\partial_c\psi^l$, 
 is actually invariant  under the $\N=1$ supersymmetry \rf{34}  by itself.
  It is the existence  of  the extended $\N=8$ supersymmetry 
 that  fixes its presence in the supermembrane action \rf{61}.

 As was already  mentioned above, 
 while the $\N=1$   supersymmetric action  \rf{41}   exists in any number of  target space-time 
 dimensions (with  index $i$ taking $D-3$ values) the $\kappa$-symmetric supermembrane action \rf{51}
 is  defined only  for  $D=4,5,7,11$  where  there is  a matching   between  the   bosonic and fermionic (target-space spinor) 
  on-shell 
 degrees of freedom \cite{Achucarro:1987nc}.  As 
  the  antisymmetric tensor 
  $U^{ijkl} $  in \rf{75}   with $i,j,...$ running $D-3$   values  
   vanishes in the  case of   $D=4$  and $D=5$   the two actions  \rf{41}  and \rf{61}  
    are, in fact, equivalent   in  these special cases (see also below).

 
Let us note  that the  $\epsilon^{abc}$ terms  in \rf{41} and \rf{61} 
 drop out  upon the double dimensional reduction 
to  2d, i.e.  to the  string  theory    case,    implying that   the corresponding actions \rf{45}  and \rf{66} should
be equivalent. 
As was shown in    \ci{Tseytlin:2025dud}, fixing the static gauge in the spinning string  action of 
\cite{Brink:1976sc,Deser:1976rb}
  and  expanding in derivatives  one gets  the $\N=1$  2d supersymmetric action for $(X^i, \psi^i)$ 
in \rf{45}. The corresponding one-loop S-matrix is then the same as  the one for the GS   action in \rf{66}. 
The relation between  the two actions  can be understood using the 2+8 split  analogous to the  3+8 one discussed above
(here $a=0,1$; cf. \rf{65}) 
\begin{align}
&\Gamma^a=\gamma^a \otimes \mathbb{I}\ ,\qquad  \Gamma^i=\gamma^2\otimes \rho^i \ , \quad \te     \gamma\Gamma_{11}=\mathbb{I}\otimes \rho^*\ ,\qquad  \mathcal{P}_\pm=\mathbb{I}\otimes p_\pm\ ,\qquad  p_\pm=\frac{1}{2}(1\pm\rho^*)\ ,  \la{77} \\
&
    \xi=\gamma^2\varepsilon\otimes\mathcal{X} ,\qquad \  \vartheta =\psi^i \otimes \eta^i\ , \qquad \  p_-\X=0 \ , \ \qquad \ \  p_+ \eta^i =0 \ . 
\la{78} \end{align}
Then  the  supersymmetry transformations  corresponding to \rf{45} and \rf{66}   become related as in \rf{71},\rf{72}, 
and as in  \rf{74} we get  
$
    \bar\theta\Gamma^a \partial_b \theta \ \to \ 
    \bar\psi'^i \gamma^a \partial_b\psi'^i$,  $\psi'^i = \sqrt{V} \psi^i, \  V= \Sigma^T\Sigma$. 
 As a result, one can establish a direct correspondence between  \rf{45}  and \rf{66}, 
 and  that matching should continue also to higher orders in expansion  in derivatives, 
 in line with the expected  equivalence of the spinning string and GS actions in a
  physical gauge. 

Let us note that one may   study similar residual world-volume supersymmetry that appears in the M2 brane action 
when it is embedded into 11d   curved background that preserves some amount of supersymmetry 
(for example, AdS$_3$   brane in AdS$^7 \times S^4$   as  in \ci{Drukker:2020swu,Beccaria:2025ahf}). 
There may   be again subtleties
 in how  the  residual 3d  supersymmetry may be related 
to the  linearly realised  one on the quadratic part of the action  (e.g. the issue of R-symmetry representation of the supersymmetry parameter, etc.). 
An example  was  mentioned in \cite{Townsend:2002wd}, where the  M2   brane is  embedded into $\mathbb R^{1,2} \times K3 \times T^4$:
the quadratic part of the action     appears to have the $\cal N$=4   supersymmetry \cite{Alvarez-Gaume:1981exv}
 while higher order terms  in the action may preserve  only $\cal N$=3   3d supersymmetry.  


\subsection{Supermembrane actions in $D=4$ and $D=5$}

 Restricting the supermembrane 
action \rf{51}   to $D=4$  and fixing the static  and $\kappa$-symmetry gauges  one  gets  the 
 action   that  has 
$\N=1$   3d supersymmetry \cite{Achucarro:1988qb}.  In $D=4$ case  there is  one transverse scalar 
 field $X= X^3$ and  the 4d   spinor $\theta$ subject to \rf{60}, 
i.e.  equal to  $\vt$  with  2  independent components   that can be identified with a 3d spinor $\psi$. 
The resulting analog of \rf{61} is then 
 \begin{align}
    \label{79}
{L} _{_{\rm 4D}}  = &\te -\frac{1}{2} \partial_a X \partial^a X +i \bar{\vartheta} \Gamma^a \partial_a \vartheta +\frac{1}{8} \partial_a X\partial_b X\partial^bX \partial^a X\no \\ &\te  -\frac{i}{2} \partial_a X\partial_b X \bar{\vartheta} \Gamma^a \partial^b \vartheta -\frac{1}{4} \bar{\vartheta} \Gamma_a \partial_b \vartheta \bar{\vartheta} \Gamma^b \partial^a \vartheta+\ldots \ .
\end{align}
The residual supersymmetry  is given by \rf{67},\rf{677}.
Using a    decomposition   similar to the one  in \rf{68}--\rf{70}  this  Lagrangian can be shown to be equivalent to  \rf{41}  for a 
 single multiplet $(X,\psi)$. 
Explicitly,    let us  set
\begin{align}\la{80}
&    \Gamma^a=\gamma^a \otimes \gamma^2,\qquad \Gamma^3= \mathbb{I}\otimes\gamma^1 ,\qquad \te 
     \quad \mathcal{P}_\pm=\mathbb{I}\otimes p_\mp,\qquad p_\mp=\frac{1}{2}(1\mp\gamma^2) \ , \\
     & \vartheta(\s) =\P_- \theta=\psi (\s)\otimes \eta \ , \qquad   \xi=\mathcal{P}_+\epsilon=\varepsilon\otimes\gamma^0\chi \ , \qquad 
     p_- \eta=0 , \ \ \  \ p_-\chi=0 \ , 
\end{align}
where $\gamma^a$ are 3d gamma matrices  defined as  in (\ref{a1})
and $\eta$ and $\chi$  are  commuting constants. After a rescaling of $\psi$ and $\ve$,  the residual 
 supersymmetry transformation \rf{67} takes the same form as  the $\N=1$  one in \rf{34} or \rf{72}
 and the relation between   the  two actions  follows from the same  argument as in \rf{74}. 

The resulting Lagrangian for $(X, \psi)$ 
may be also
 interpreted (and  expressed to all orders in the $\N=1$ superfield  form) 
   following    the nonlinear realization approach  
\cite{Volkov:1973ix,Hughes:1986dn,Hughes:1986fa}, 
  as  describing an   effective theory of  Goldstone modes $X$ and $\psi$  associated with
  the spontaneous breaking of target-space supersymmetry  
  \cite{Ivanov:1999fwa,Ivanov:2001gd,Bellucci:2000kc,Bellucci:2013pba}
  (for similar non-linear  actions  involving vectors  and spinors like D-brane   actions 
  see  \ci{Bergshoeff:2013pia} and refs. there).


In   the $D=5$ case  we get  two real   or one complex  $(\XX=X^3+ i X^4,\, \pps=\psi^3 + i\psi^4)$  3d multiplet
as  representing the physical   degrees of freedom. 
The resulting actions     \rf{41}  or  \rf{61}  are  then  equivalent     and again may be also
 interpreted  as  describing an   effective theory of   Goldstone modes $X^i$ and $\psi^i$  associated with the partial 
 spontaneous breaking of target-space supersymmetry.

The  static gauge fixing in   the $D=5$  bosonic  membrane action
breaks the  Poincar\'e symmetry $ISO(1,4)$ to $ISO(1,2)\times SO(2)$ and similarly for the  supermembrane case
where the $\N=1$     supersymmetry in $D=5$  is spontaneously broken to $\N=2$  supersymmetry in 3d. 
 The resulting $\N=2$  supersymmetric  action of  the $D=5$ supermembrane    may  be written as  \cite{Bellucci:2013mha}
\begin{align}
   \mathcal{S}_{_{\text{5D}}}=- T  \int d^3\sigma\;  \mathcal{E}\,  \Big[1 &+ \sqrt{(1+\nabla_a \XX \nabla ^a \bar \XX)^2
  -(\nabla_a \XX   \nabla ^a  \XX)(\nabla_b \bar  \XX \nabla ^b \bar  \XX)} \nonumber\\
   &\qquad
   +\tfrac{i}{\sqrt{2}}\epsilon^{abc}\nabla_a \XX \nabla_b\bar \XX (\bar\Psi\nabla_c \Psi+\nabla_c \bar\Psi \Psi) \Big]\ , \la{81}\\
   \mathcal{E} \equiv  \det  \mathcal{E}_a\,^b \ , \qquad  &\ \ \
     \mathcal{E}_a\,^b=\delta_a^b-\tfrac{i}{\sqrt{2}}(\bar \Psi \gamma^b \partial_a \Psi-\partial_a\bar\Psi\gamma^b\Psi)\ ,  \la{82}\\
     \nabla_a\equiv (\mathcal{E}^{-1})_a\,^b\, \partial_b &=\partial_a+\tfrac{i}{\sqrt{2}}\big(\bar \Psi \gamma^b\nabla_a\Psi-\nabla_a\bar\Psi \gamma^b\Psi\big)\partial_b \ ,  \la{83}
\end{align}
\iffa 
\begin{align}
   \mathcal{S}_{_{\text{5D}}}=- T  \int d^3\sigma\;  \mathcal{E}\,  \Big[1 &+ \sqrt{(1+\nabla_a \XX \nabla ^a \bar \XX)^2
  -(\nabla_a \XX   \nabla ^a  \XX)(\nabla_b \bar  \XX \nabla ^b \bar  \XX)} \nonumber\\
   &\qquad 
   -\tfrac{1}{\sqrt{2}}\epsilon^{abc}\nabla_a \XX \nabla_b\bar \XX (\bar\Psi\nabla_c \Psi-\nabla_c \bar\Psi \Psi) \Big]\ , \la{81}\\
   \mathcal{E} \equiv  \det  \mathcal{E}_a\,^b \ , \qquad  &\ \ \ 
     \mathcal{E}_a\,^b=\delta_a^b+\tfrac{1}{\sqrt{2}}(\bar \Psi \gamma^b \partial_a \Psi+\partial_a\bar\Psi\gamma^b\Psi)\ ,  \la{82}\\
     \nabla_a\equiv (\mathcal{E}^{-1})_a\,^b\, \partial_b &=\partial_a+\sqrt{2}\big(\bar \Psi \gamma^b\nabla_a\Psi+\nabla_a\bar\Psi \gamma^b\Psi\big)\partial_b \ ,  \la{83}
\end{align}
\fi
where  $\bar \XX$ is the complex conjugate of $\XX $ and  
$\bar \Psi=\Psi^\dagger \gamma^0$. 
Written 
in  terms of  the  real fields $(X^i, \psi^i)$  and expanded in derivatives  the  Lagrangian in \rf{81}   becomes 
\begin{align}
    \mathcal{L}_{_{\text{5D}}}=&\te- \partial_a X^i\partial^a X^i +2\sqrt{2}i\bar\psi^i \gamma^a\partial_a\psi^i-\frac{1}{2}\partial_a X^i\partial^a X^i \partial_b X^j \partial^b X^j+\partial_a X^i \partial_b X^i \partial^a X^j \partial^b X^j\la{84}\\
    &\te -2\sqrt{2}i\partial_a X^i \partial_b X^i\bar \psi^j \gamma^a\partial^b\psi^j-2\sqrt{2}i\epsilon ^{abc}\partial_a X^i \partial_b X^j \bar \psi^i \partial_c \psi^j   \te -2\bar\psi^i \gamma_a\partial_b\psi^i \bar \psi^j \gamma^b\partial^a \psi^j+... \ , \no
\end{align}
where  we ignore a term proportional to the leading-order fermionic equations of motion. 
Upon  the  rescaling $X^i\rightarrow 2^{-1/2}X^i$ and $\psi^i\rightarrow 2^{-3/4}\psi^i $  this  takes the same 
 form as   (\ref{41}).

\section{One-loop  S-matrices} 

Let   us now  compare the one-loop  $2\rightarrow 2$ bosonic scattering amplitudes
that correspond  to the $\N=1$  supersymmetric Lagrangian in \rf{41} and the   supermembrane Lagrangian   in \rf{61}.
The one-loop S-matrix for the  latter was already found in \ci{Seibold:2023zkz}. 

In general, the  $2\rightarrow 2$   scalar amplitude  has the structure 
\begin{align}
  & \ \ \  \mathcal{M}_{i j, k l}=A(s,t,u)  \delta_{i j} \delta_{k l}+B(s,t,u) \delta_{i k} \delta_{j l}+C(s,t,u) \delta_{i l} \delta_{j k}\ ,\label{91}\\
  &
     B(s,t,u)=A(t,s,u)\ ,\qquad \quad C (s,t,u)=A(u,t,s)\ , \qquad  A^{(0)} =-\tfrac{1}{2} ut \ ,    \qquad s+t+u =0  \ , \la{911}
\end{align}
where  $ A^{(0)}$  is the  tree-level amplitude    corresponding to the quartic terms in \rf{32}.
The bosonic loop contribution to the  one-loop amplitude is  given by  \ci{Seibold:2024oyr,Seibold:2023zkz}\foot{The one-loop amplitude in 3d 
 is UV finite in dimensional regularization.  Note that we  ignore the factor $T^{-1}$ of  membrane tension in the normalization of the amplitude.}
\begin{align}
    &A^{(1)}_b=\te \tfrac{1}{256} \big[(-s)^{3 / 2}\big[(\frac{3 }{32}\hat{D} -1) s^2-\frac{1}{4}\hat{D}  t u\big]+(-t)^{3 / 2} t(3 t+2 s)+(-u)^{3 / 2} u(3 u+2 s)\big] \ , \la{92}
\end{align}
where $ \hat D=D-3$ is the number of  components of $X^i$, i.e. the   number of transverse directions. 

To find the  fermionic loop contribution we need  the vertex   corresponding the  $\del X \del X \bar \psi \del \psi$ terms in \rf{41}:
\begin{align}\la{09}
 \mathcal{V}_{ijkl} =\te  -\frac{1}{2}\big[(\gamma\cdot p_1)(p_2\cdot p_4)+(\gamma\cdot p_2)(p_1\cdot p_4)\big]\delta_{ij}\delta_{kl}-\frac{1}{2}\epsilon^{abc}\, p_{1a} p_{2b} p_{4c}\big(\delta_{ik}\delta_{jl}-\delta_{il}\delta_{jk}\big)\ . 
\end{align}
The  fermionic loop  contribution to  the scalar scattering amplitude  containing two such vertices   is 
\begin{align}
\label{93}
    A^{(1)}_f
    &=\tfrac{1}{2048}n_f\Big[\tfrac{1}{8}\hat D
   (-s)^{3 / 2}\big(s^2-8 t u\big)  + (-t)^{3 / 2}t(t+2s)+(-u)^{3 / 2}u(u+2s)\Big]\ . 
\end{align}
where $n_f=2$ is the number of  3d spinor  components of $\psi^i$.
 
Starting with  the  BST Lagrangian \rf{61}   one finds  for the fermionic loop contribution  \ci{Seibold:2024oyr}
\begin{align}
\label{94}
    A^{(1)}_{ {\rm M},f}
    &=\tfrac{1}{1024\times 16 }{\rm n}_{\rm F} \Big[
   (-s)^{3 / 2}\big(s^2-8 t u\big)  + 4(-t)^{3 / 2}t(t+2s)+4(-u)^{3 / 2}u(u+2s)\Big]\ , \\
   & \qquad {\rm n}_{\rm F} = n_f  \hat D = 2 \hat D \ , \la{95} 
\end{align}
where ${\rm n}_{\rm F}$   is the number   target space spinor  components 
(for $D= 4,5,7,11$ where the number of physical  bosonic and  fermionic  3d degrees of freedom match). 

We observe that   for $D=11$, i.e. $\hat D=8$,  the  expressions  in \rf{93}   and \rf{94}  differ  by  extra  factors of 4  in the last 2 terms in \rf{94}. 
 This difference is  a consequence of the  inequivalence of the 
 $\eps ^{abc}$ terms    in \rf{41} and in \rf{61} discussed above  (cf. \rf{75}). 
 Indeed, the contribution of the  $-\frac{i}{4} \epsilon^{a b c} \partial_a X^i \partial_b X^j \bar{\vartheta} \Gamma_{i j} \partial_c \vartheta $ 
  term  in \rf{61} to the one-loop S-matrix is 4 times  that of the
 $-\frac{i}{2}\epsilon^{abc}\partial_{a}X^{i}\partial_{b}X^{j}\bar{\psi}^{i}\partial_{c}\psi^{j} $ term in \rf{41}.

 At the same time,  the expressions in \rf{93}  and  \rf{94} match for $D=5$, i.e. for  $\hat D=2$.
 The   two amplitudes also  match  in the case of $D=4$ where   there is  only  one  boson  (and so 
 the $\epsilon^{abc}$ term in \rf{31} is absent).  For $D=4$    one gets from \rf{41} 
 for the total  fermionic loop contribution to the  amplitude (cf. \rf{83}) 
 \begin{align}\la{98}
  \no   \mathcal{M}^{(1)}_f=   &  A^{(1)}_f +   B^{(1)}_f +   C^{(1)}_f \\ =  &
    - \tfrac{1}{16384}n_f \Big[(-s)^{5 / 2} s+(-t)^{5 / 2} t+(-u)^{5 / 2} u-8 s t u(\sqrt{-s}+\sqrt{-t}+\sqrt{-u}\, )\Big] \ . 
\end{align}
 This  agrees  with the $D=4$  supermembrane result \ci{Seibold:2024oyr} 
 that follows from \rf{94}, i.e.  with  $ \mathcal{M}^{(1)}_{{\rm M},f}=     A^{(1)}_{{\rm M},f} +   B^{(1)}_{{\rm M},f} +   C^{(1)}_{{\rm M},f}$.

 The total one-loop amplitude  in $D$ dimensions  is obtained by summing together  the  bosonic  and  the fermionic   loop contributions. 
  Adding together  \rf{92}   and \rf{93}  we get   the following expression corresponding to 
  the $\N=1$ 3d supersymmetric action in \rf{41} 
 \begin{align}
\label{101}
{A}^{(1)}=&\te 
(-s)^{3/2}\left[
\frac{(6+n_f)\hat D - 64}{16384}\, s^2
- \frac{(2+n_f)\hat D}{2048}\, t u
\right]\no \\
&\te 
+ (-t)^{3/2}t\left[\frac{8+n_f}{1024}\, s +
\frac{24+n_f}{2048}\, t 
\right]
+ (-u)^{3/2}u\left[ \frac{8+n_f}{1024}\, s +
\frac{24+n_f}{2048}\, u
\right] \ . 
\end{align}
Specifying to   $n_f=2$ and $\hat D=8$ this  gives
 \be \la{1010}
{A}^{(1)}\Big|_{D=11} =\te -\frac{1}{64}\Big[
(-s)^{3/2}  t u 
- {1\ov 16}  (-t)^{3/2} t\, (10s + 13t)
- {1\ov 16}  (-u)^{3/2} u\, (10s + 13u)
\Big] \, .
\ee
 At the same time,  from 
the  supermembrane    action   \rf{61} we get  for 
the sum of \rf{92}  and \rf{94} (with  ${\rm n}_{\rm F} = n_f  \hat D$)  
\begin{align}
 {A}^{(1)}_{\rm M}= &\te (-s)^{3/2}\left[
\frac{(6+n_f)\hat D - 64}{16384}\, s^2
-\frac{(2+n_f)\hat D}{2048}\, t u
\right]
\\
& \te + (-t)^{3/2}t\big[\frac{8+\ha n_f\hat D}{1024}\, s +
\frac{24+\ha n_f\hat D}{2048}\, t
\big]
+
(-u)^{3/2}u\big[\frac{8+\ha n_f\hat D}{1024}\, s + 
\frac{24+\ha n_f\hat D}{2048}\, u
\big]\, . \la{99}
\end{align}
Compared to \rf{101} the difference is in coefficients in the second line. 

As already discussed above, the expressions  in \rf{101}  and \rf{99}  should agree for $D=5$, i.e. for 
$n_f=2, \ \hat D=2$,  where  they indeed reduce to 
\be   \te {A}^{(1)}\Big|_{D=5}={A}^{(1)}_{\rm M}\Big|_{D=5}=
\frac{1}{1024}\Big[
(-s)^{3/2}\big(-3 s^2 - 4 t u\big)
+ (-t)^{3/2}t\,(13 t + 10 s)
+ (-u)^{3/2}u\,(13 u + 10 s)
\Big]\, .\la{103}
\ee 
For $D=11$   case, i.e. $n_f=2, \ \hat D =8$,  the  supermembrane expression in \rf{99}   simplifies to  (cf. \rf{101})
 \ci{Seibold:2024oyr}
 \be 
 \la{100}
  {A}^{(1)}_{\rm M}\Big|_{D=11}=- \tfrac{1}{64 } \Big[
   (-s)^{3 / 2}  + (-t)^{3 / 2}+(-u)^{3 / 2}\Big] tu\, .
 \ee
  

\

\section*{Acknowledgements}
AT is grateful to M. Beccaria and S. Kurlyand for helpful discussions and comments on the draft. He also thanks R. Kallosh for useful remarks.
ZW would like to thank K. Mkrtchyan, D. Waldram, and D. Zhong for discussions.
This work was supported by the STFC grant ST/T000791/1.

We dedicate this paper to our late colleague and advisor Kellogg Stelle (11.03.1948–23.10.2025), who made  important contributions to the development of the supermembrane theory
\ci{Duff:1987cs,Duff:1987bx,Gandhi:1988sh,Pope:1988dv,Pope:1989hf,Pope:1989cka,Achucarro:1989dd,Duff:1990xz}.
Kelly played a key role in bringing AAT to Imperial College in 1992. He also  generously accepted ZW as his PhD student in 2023.

\newpage 

\appendix
\section{Spinor relations\la{apA}}

For 3d Dirac matrices   and   2-component Majorana  spinors   satisfy  the following relations 
\ba
 &   
    \gamma^{0}=\Big(\begin{array}{cc}
0 & -1\\
1 & 0
\end{array}\Big),\quad\gamma^{1}=\Big(\begin{array}{ll}
0 & 1\\
1 & 0
\end{array}\Big),\quad\gamma^{2}=\Big(\begin{array}{cc}
1 & 0\\
0 & -1
\end{array}\Big)\ , \la{a1} \\
&\te \gamma^{ab}=\epsilon^{abc}\gamma_{c},\qquad\gamma^{a}=-\ha \epsilon^{abc}\gamma_{bc},\qquad\gamma^{abc}=\epsilon^{abc}\mathbb{I},\qquad\epsilon_{012}=1,\qquad\epsilon^{012}=-1\ , \label{a0}
\\
&    \bar{\psi}\chi=\bar{\chi}\psi,\qquad\bar{\psi}\gamma^{a}\chi=-\bar{\chi}\gamma^{a}\psi,\qquad\bar{\psi}\gamma^{a}\psi=0,\qquad\bar{\psi}\gamma^{a}\partial_{a}\psi=-\partial_{a}\bar{\psi}\gamma^{a}\psi\ ,\qquad \ \ \ 
 \bar{\psi}=\psi^{T}\gamma^{0}\ , \la{a4} \\
    \label{a2}
&  \te    (\bar{\varphi} \chi)(\bar{\psi} \lambda)=-\frac{1}{2}(\bar{\psi} \chi)(\bar{\varphi} \lambda)-\frac{1}{2}\big(\bar{\psi} \gamma^a \chi\big)\big(\bar{\varphi} \gamma_a \lambda\big) \,
 . 
\end{align} 
For 11D   Dirac matrices  and Majorana  spinor $\theta$   with $\P_\pm \equiv \ha (1 + \Gamma^{012})$ as in \rf{56},\rf{57} one has 
($a=0,1,2; \ i=3, ...,10$)
\begin{equation}
\gamma\Gamma^i=-\Gamma^i \gamma,\qquad \gamma\Gamma^a=\Gamma^a \gamma,\qquad \gamma\mathcal{P}_\pm=\pm\mathcal{P}_\pm,\qquad \overline{\mathcal{P}_\pm\theta}=\bar \theta \mathcal{P}_\pm, \qquad \gamma\equiv \Gamma^{012}.
\end{equation}
\iffa 
Some properties used in the case of 2d reduction to  GS string are 
\begin{equation}
\gamma\Gamma^{11} \Gamma^i=-\Gamma^i \gamma\Gamma^{11}, \quad \gamma \Gamma^{11} \Gamma^a=\Gamma^a \gamma \Gamma^{11}, \quad \gamma \Gamma^{11}\mathcal{P}_{ \pm}= \pm \mathcal{P}_{ \pm}, \quad \overline{\mathcal{P}_{ \pm} \theta}=\bar{\theta} \mathcal{P}_{\mp}, \quad \gamma=\Gamma^{01}\  .
\end{equation}
\fi 
With $\vt= \P_-\theta$  as in \rf{58} one has 
\begin{equation}
    \Gamma^{abc}\vartheta=-\epsilon^{abc}\gamma\vartheta =-\epsilon^{abc}\gamma\mathcal{P}_-\vartheta=\epsilon^{abc}\vartheta \ , 
\end{equation}
so that acting   on $\vt$ the matrices $\Gamma^a$  satisfy the same relations as the   $\gamma^a$ ones. 

\section{$\N=1$ supersymmetry of the $U^{ijkl}$ term  in the BST action \la{apC}   }

The additional term containing the constant antisymmetric tensor  $U^{ijkl}$
 that appears in \rf{75}, when combined with the expression \rf{700} for the BST target-space  spinor 
 $\vt$ in terms of 3d   spinors $\psi^i$
gives rise to the following extra contribution to the supermembrane Lagrangian in \rf{61}:
\begin{align}
&\te 
\Delta L=-\frac{i}{4} \epsilon^{a b c} \partial_a X^i \partial_b X^j \bar{\vartheta} \Gamma_{i j} \partial_c \vartheta \ \ \to \ \ \ 
\Delta \L=-\frac{i}{4} Y \ , \ \ \ \ \  Y\equiv  \epsilon^{a b c}  U^{ijkl} \partial_a X^i \partial_b X^j  \bar\psi^k\partial_c\psi^l\ .\la{c1}
\end{align}
Here  we will show that this term is separately invariant under the linear $\N=1$   supersymmetry \rf{34} or \rf{72}.
 Integrating by parts  and 
ignoring a total derivative term  we    can write the supersymmetry variation of $Y$ as 
\be 
\delta Y=
-\epsilon^{abc}U^{ijkl}\partial_{a}X^{i}\partial_{b}X^{j}\partial_{d}X^{k}\, \bar{\varepsilon}\gamma^{d}\partial_{c}\psi^{l}-2i\epsilon^{abc}U^{ijkl}\partial_{a}X^{i} \bar{\varepsilon}\psi^{j} \, \partial_{b}\bar{\psi}^{k}\partial_{c}\psi^{l}
\ . \label{c2}
\ee
In the  first  term in \rf{c2}  the antisymmetry in $i,j,k$ implies that $\partial_{a}X^{i}\partial_{b}X^{j}\partial_{d}X^{k} \sim \ve_{abd}$
and  as a result   it is the same as
 $\sim \epsilon^{abd}U^{ijkl}\partial_{a}X^{i}\partial_{b}X^{j}\partial_{d}X^{k}\, \bar{\varepsilon}\gamma^{c}\partial_{c}\psi^{l}
$
which is 
proportional to  the leading-order equation motion following from \rf{33}. Thus it is an on-shell  superinvariant (equivalently, this 
  this term  can be  absorbed into a deformation  of the $\psi^i$  transformation law). 

The   second term in \rf{c2}   can be  rearranged     using the Fierz identity  \rf{a2} and integration by parts   
\begin{align}
-2i\epsilon^{abc}U^{ijkl}\partial_{a}X^{i} \bar{\varepsilon}\psi^{j} \, \partial_{b}\bar{\psi}^{k}\partial_{c}\psi^{l}
  \to  -i\epsilon^{abc}U^{ijkl}\partial_{a}X^{i}\bar{\varepsilon}\psi^{k}\partial_{b} \bar{\psi}^{j}\partial_{c}\psi^{l}
+i\epsilon^{abc}U^{ijkl}\partial_{a}X^{i}\bar{\varepsilon}\gamma^{e}\psi^{k}\partial_{b}\bar{\psi}^{j}\gamma_{e}\partial_{c}\psi^{l}.\label{c3}
\end{align}
The last term in (\ref{c3}) vanishes    due  to  the Majorana   relation in \rf{a4} 
($\partial_{b}\bar{\psi}^{j}\gamma_{e}\partial_{c}\psi^{l}=-\partial_{c}\bar{\psi}^{l}\gamma_{e}\partial_{b}\psi^{j}$)
 and   the antisymmetry in  $j,l$ and also in  $b,c$.   
The first term in  the r.h.s. of  (\ref{c3})  is the same   as the one on the  l.h.s.  (\ref{c3}) up to  the  -2   coefficient 
and thus   should also vanish.

The on-shell supersymmetry  of $Y$ in \rf{c1} can be  seen also directly  from  its   representation as 
the last component   of the following   superfield (cf. \rf{35}--\rf{40}) 
\be \la{c4}
Y= - 8i  
 U^{ijkl} \Big[\DD^\alpha \Phi^i \DD^\beta \DD^\gamma \Phi^j \DD_\beta \Phi^k \DD_\gamma \DD_\alpha \Phi^l\Big]_{ \bar \theta \theta} + ...
\ , \ee 
where dots stand for a  term vanishing on the free  fermionic equation of motion.


\section{One-loop S-matrix  for   $\N=1$  3d   supersymmetric action
\label{appB}}

Here   we shall consider  the following  generalization of the  Lagrangian  in \rf{33},\rf{41} 
where  we combine the two superinvariants \rf{39}  and \rf{40} with an arbitrary coefficient $\zeta$, i.e. 
\ba
 \mathcal{L} =& -\tfrac{1}{2}\partial_a X^i \partial^a X^i +i\bar \psi^i \gamma^a \partial_a\psi^i - \tfrac{1}{8} \mathcal{I}_1'+ ...\ ,\la{b0} \\
\mathcal{I}_1' \equiv &\te \mathcal{I}_2+\frac{1}{2}\z\big(\mathcal{I}_2-\mathcal{I}_1\big)
=\partial_a X^i \partial^a X^i \partial_b X^j \partial^b X^j+\zeta \partial_a X^i \partial_b X^i \partial^a X^j \partial^b X^j\nonumber\\
&\ \ +2 i\zeta\, \partial_a X^i \partial_b X^i\bar{\psi}^j \gamma^b \partial^a \psi^j+2i\zeta\, \partial_a X^i \partial_b X^j\bar{\psi}^i \gamma^b \partial^a \psi^j+ 4i \partial_a X^i \partial_b X^j\bar{\psi}^i \gamma^a \partial^b \psi^j\nonumber\\
&\ \  +(4+\zeta)\bar{\psi}^j \gamma_a \partial_b \psi^j\bar{\psi}^i \gamma^b \partial^a \psi^i-2(2+\zeta)\bar{\psi}^j \gamma_b \partial_a \psi^j\bar{\psi}^i \gamma^b \partial^a \psi^i.\label{b1}
\end{align}
The Lagrangian \rf{b0} reduces back to \rf{41}  for $\z=-2$. 
We  shall  consider the   scattering of massless  world-volume   scalars $X^i$  with 
the  Mandelstam variables  defined as  
\begin{align}
    s=-(p_1+p_2)^2\ , \qquad 
     t=-(p_1-p_3)^2 \ , \qquad 
      u=-(p_1-p_4)^2\ , \qquad \ \  s+t + u=0 \ .
\end{align}
The  tree-level amplitude in  \rf{91}  following from \rf{b1}  is then   given by (cf. \rf{911}) 
 \begin{align}\te 
A^{(0)} (s,t,u) =-\frac{1}{8}\big[(2+\zeta) s^2+2 \zeta u(s+u) \big]  \ . 
\end{align}
The  bosonic loop   contribution to the  one-loop amplitude is  found to be\foot{The  amplitudes are   computed using dimension regularization with $d=3-2\varepsilon$ as in   \cite{Seibold:2024oyr}. }
\begin{align}
     A^{(1)}_b (s,t,u) =&\te \tfrac{(-s)^{3 / 2}}{65536}\Big[-16 t u \zeta \big[4+(2+\hat{D}) \zeta\big]+2 s^2\big[\hat{D}\left(128+96 \zeta+19 \z^2\right)+2\left(96+214 \zeta+67 \zeta^2\right)\big]\Big]
\nonumber\\
    &\te +\frac{(-t)^{3/2}}{65536}\Big[8 s^2(2+\zeta)^2+8 s t\big(20-12 \zeta+5 \zeta^2\big)+t^2\big(140+204 \zeta+259 \zeta^2\big)\Big]\nonumber\\
    &\te +\frac{(-u)^{3/2}}{65536}\Big[{u}^2\big(140+204 \zeta+259 \zeta^2\big)+32 {s}{u}(2-\z)^2- 8s{t}(2+\zeta)^2\big)\Big] \ . \label{b2}
\end{align}
This reduces to \rf{92} for $\z=-2$. 

The fermionic loop   contribution to  one-loop $s$-channel amplitude  in the theory  defined by \rf{b0},\rf{b1}
 can be written as 
\begin{align}
 & \mathcal{M}^{(1)}_{f,s}=\tfrac{1}{2i}\int^1_0 dx \int\frac{d^d l}{(4\pi)^d}\frac{N_{f,s}}{(l^2+\Delta)^2}\ ,   \label{b3}
 \qquad \ \ \   l=p-x(p_1+p_2),\qquad \Delta =-x(1-x)s\,, \\
& N_{f,s}  =\mathcal{V}_{ijmn}(p_{1},p_{2},p,q)\, \mathcal{V}_{klnm}(-p_{3},-p_{4},-q,-p)\, {\rm Tr}\big[(-i p) \cdot \gamma(i q) \cdot \gamma\big]\  , 
\end{align}
where $\mathcal{V}_{ijmn}$ is the  2-boson -- 2-fermion vertex in \rf{b1}.    If $N_{f,s}$  is  decomposed as 
\begin{align}
    N_{f, s}
    =N_0+N_2 l^2+N_{a b} l^a l^b+N_4\big(l^2\big)^2+N_{a b c e}l^a l^b l^c l^e+M_{a b} l^2 l^a l^b\ , 
\end{align}
then the  integral in (\ref{b3}) can   be written  as 
\begin{align}
& \tfrac{1}{2i}\int^1_0 dx \int \frac{d^d l}{(4\pi)^d}\frac{N_{f,s}}{(l^2+\Delta)^2} =\tfrac{1}{16 \pi} \int_0^1 d x\Big[N_0\Delta^{-\frac{1}{2}}-3 N_2\Delta^{\frac{1}{2}}-N_{a b} \eta^{a b}\Delta^{\frac{1}{2}}\nonumber \\
&\qquad \qquad \qquad\qquad\qquad  \te +5 N_4\Delta^{\frac{3}{2}}+\frac{1}{3} N_{a b c e}\Delta^{\frac{3}{2}}\big(\eta^{a c} \eta^{b e}+\eta^{a b} \eta^{c e}+\eta^{a e} \eta^{b c}\big)+\frac{5}{3} M_{a b}\Delta^{\frac{3}{2}}\big(\eta^{a b}\big)\Big]\ . 
\end{align}
The resulting  fermionic  loop contribution  to the  amplitude  $A^{(1)}$ is found to be
\begin{align}
A^{(1)} = & A_{f, s}^{(1)}+ A_{f, t}^{(1)} + A_{f, u}^{(1)}   \te =\frac{n_f \zeta[(2+\hat{D}) \zeta+4]}{4 \times 64 \times 256}(-s)^{3 / 2}\big(s^2-8 t u\big)\no  \\
 & \te +\frac{n_f}{8 \times 64 \times 256} (-t)^{3 / 2}\big[t^2\big(20-12 \zeta+5 \zeta^2\big)-8 s\big(s(-2+\zeta)^2+2 u\big(4+\zeta^2\big)\big)\big]\no \\
 & \te+\tfrac{n_f}{8 \times 64 \times 256}(-u)^{3 / 2}\left[u^2\left(12-20 \zeta+3 \zeta^2\right)-8 t\left(t(-2+\zeta)^2-8 s \zeta\right)\right]
 \ . \la{b10}
\end{align}
This reduces to \rf{93} for $\z=-2$. 

Similar  computation can be carried out   for the corresponding  reduction  of \rf{b0} to 2 dimensions.
 In this case  for the  $t=0, u=-s$ choice of the massless 2d  kinematics
one finds that  the tree-level amplitudes are 
\begin{equation}
  \te    A^{(0)}=-\frac{1}{8}(2+\zeta)s^2, \qquad
 B^{(0)}=-\frac{1}{4}\z\,s^2,  \qquad
 C^{(0)}=-\frac{1}{8}(2+\zeta)s^2 \ . 
\end{equation}
It is only for  $\zeta=-2$  which corresponds to the   combination \rf{32} of the bosonic 4-derivative terms in \rf{b1}  that 
 comes out of the expansion of the Nambu action 
    the transmission amplitude $B^{(0)}$ is  the only one that is non-zero, so that  the Yang-Baxter equation is satisfied.
Similarly, the  sum of the  bosonic and fermionic  loop   contributions gives the expression 
 consistent with integrability only for $\z=-2$
(cf. \cite{Seibold:2024oyr}). 

\newpage 

{\small 
\bibliographystyle{JHEP-v2.9}
\small
\bibliography{biblio2.bib}
}
\end{document}